\documentclass[nofootinbib,aps,prd,preprint,superscriptaddress]{revtex4}
\usepackage{epsfig,epstopdf}
\usepackage[utf8x]{inputenc}
\usepackage{amsmath,graphicx}
\usepackage{color}

\newcommand{\eps}{\varepsilon}
\newcommand{\ga}{\gamma}

\newcommand{\nn}{\nonumber}

\newcommand{\bn}{{\bar n}}

\newcommand{\pslash}{{\not \!p}}
\newcommand{\kslash}{{\not \!k}}
\newcommand{\nslash}{{\not \!n}}
\begin{document}
\baselineskip 3.0ex

\vspace*{18pt}

\title{Singular and Regular Gauges in Soft Collinear Effective Theory: The Introduction of the New Wilson Line  T}

\author{Ahmad Idilbi}\email{ahmadidilbi@gmail.com}
\author{Ignazio Scimemi}\email{ignazios@fis.ucm.es}

\affiliation{Departamento de F\'isica Te\'orica II,
Universidad Complutense de Madrid (UCM),
28040 Madrid, Spain}

\begin{abstract}
\baselineskip 3.0ex  \vspace{0.5cm}
Gauge invariance in soft-collinear effective theory (SCET) is discussed in regular (covariant) and singular (light-cone) gauges.
It is argued that SCET, as it stands, is not capable to define in a gauge invariant way certain non-perturbative matrix elements
that are an integral part of many factorization theorems.
Those matrix elements involve two quark or gluon fields separated not only in light-cone direction but also in the transverse one.
This observation limits the range of applicability of SCET. To remedy this we argue that one needs to introduce a new Wilson line as part of SCET
formalism, that we call $T$. This Wilson line depends only on the transverse component of the gluon field. As such it is a new feature to the SCET
formalism and
it guarantees gauge invariance of the non-perturbative matrix elements in both classes of gauges.
 %With the presence of this Wilson line one is enforced to introduce new Lagrangian of soft and collinear interactions.
%This new Lagrangian will enable us to establish factorization theorems for high-energy processes and define the non-perturbative matrix elements
%in a gauge invariant way. A feature which is missing from the familiar SCET Lagrangian.

\end{abstract}
\maketitle
%\section{Introduction}
In the era of the Large Hadron Collider (LHC) where two hadron beams collide, factorization theorems of the various cross sections of phenomenological
interest are the basis of any phenomenological analysis. In recent years, Soft-Collinear Effective Theory
(SCET)~\cite{SCET1,SCETf,Bauer:2002nz} has emerged as the most effective framework within which such factorization theorems
can be established in a rather coherent manner. The success of SCET to factorize high-energy processes
(and thereby allowing resummation of large logarithms) rests mainly on two factors. The first is the de-coupling of the soft and collinear modes,
responsible of the infra-red (IR) divergences in perturbative QCD, at the level of the SCET Lagrangian and operators. Identifying those IR divergences,
essentially, is the most important step in establishing factorization theorems since it allows for an operator definition of the non-perturbative
hadronic matrix elements.
The second factor is the \emph{gauge invariant} building-blocks of SCET which render the previous operator definitions gauge invariant as they should
be.
 In SCET there are two fundamental quantities with which hadronic matrix elements can be defined.
 In the quark-gluon collinear sector we have $W_{\bn}^{\dagger} \xi_{\bn}$ and $W_{\bn}G^{\mu\nu}W_{\bn}$ for gluon-gluon collinear interactions.
 It is well-known that those quantities are gauge invariant (under collinear and soft gauge transformations) as long as the gauge transformation is unity at infinity (in other words $\alpha^a(x^-=\infty)=0$ where $\alpha^a(x)$ is the gauge transformation matrix.) This is the case in covariant gauges.

 The covariant gauge is ``regular'' in the sense that all components of the gluon field $A^\mu$ vanish at light-cone infinity while for ``singular''
 gauge, certain components do not.
Actually one can show that  in QED in light-cone gauge we have the following \cite{jack},
 \begin{eqnarray}
 A^+=A^-=0, \,\,\,\,\,A_\perp=-\frac{e}{2\pi}\theta(x^-)\nabla \ln \mu r_\perp \ ,
 \end{eqnarray}
where $e$ is the electric charge, $x^-$ is a light-cone coordinate and $r_\perp$ is the distance in the transverse direction.
$\mu$ is an arbitrary mass scale needed for dimensional purposes.

The fact that in light-cone gauge $A_\perp$ does not vanish at $x^-=\infty$ (unlike the case in covariant gauge) will have a rather very important
impact on the ``gauge invariant'' SCET matrix elements as we shall see below.
The basic observation is that in singular gauges a gauge transformation at infinity in light-cone direction, can be applied on the quark field which has to be, as usual, compensated for by a Wilson line.
 The result of this simple observation is that in light cone gauge one is enforced to introduce a new Wilson lines,
 built only from the transverse components $A^\mu_\perp$. We will  call  these Wilson lines $T,\ T^\dagger$
as they contain only the transverse components of the gluon field.
  The $T$-Wilson line involves only the transverse component of the gluon field. It is a new feature to be added to the definition of SCET matrix elements.

There are two main results for the introduction of the $T$-Wilson line.
The first is that SCET will not be altered once we work in covariant gauges since $T=1$ in that case.
However, and more importantly, it will allow for gauge invariant definitions of the non-perturbative matrix elements when the two quark or
gluon fields are separated not only in the $x^-$ direction but also in the transverse ones as well.
 As an example, consider the transverse-momentum-dependent parton distribution function (TMDPDF).
This function enters the factorization theorem of the semi-inclusive deep-inelastic scattering (SIDIS) processes,
see Ref.~\cite{ji,cs}.
The SCET version of the TMDPDF, built only from the collinear quark field and the collinear Wilson line, is not gauge invariant.
As such the SCET TMDPDF is not really a genuine physical quantity. However when the contribution from the $T$-Wilson line is taken into account it becomes gauge invariant
in both classes of gauges: regular and singular ones.
More discussion on this will be given below and in a forthcoming paper~\cite{future}.

An  important piece of discussion in light-cone gauge concerns the type of prescription to be used in the gluon propagator.
 In Ref.~\cite{Bassetto:1984dq} it was  shown that  the canonical quantization of QCD in light-cone gauge is consistent only with the
Mandelstam-Leibbrandt (ML) prescription.  This prescription is the only one  which respects causality, unitarity, QCD power counting and allows
for Wick rotations in loop integrals. Clearly, this prescription must be taken as the reference prescription also in SCET.
The use of  a particular prescription is  related to  the regularization of infrared (IR) divergences, as  the prescription regulates
essentially infrared  singularities. Thus the use  of prescriptions  different from  ML's can   be justified  in some cases with a
compatible regularization   of IR  divergences. We discuss  this issue in more detail below.

\section{The SCET jet in Feynman gauge  and in Light-Cone gauge with ML prescription}

Let us consider the fundamental quantity in SCET $\langle 0| W_\bn^\dagger \xi_\bn| q_\bn\rangle$ which describes an incoming parton moving along
the $\bn=\frac{1}{\sqrt 2}(1,0,0)$ direction (our notation is $(+,-,\perp)$ for light-cone coordinates). We will demonstrate below, at the one-loop
level, that this quantity, with the zero-bin subtracted \cite {Manohar:2006nz} (i.e. the pure collinear jet) is not gauge invariant. It will attain different results in Feynman gauge and in light-cone gauge.
 It should be mentioned however that the last statement is true depending on the prescription used to regularize the spurious ``light-cone singularity''
originating from the $1/k^+$ term in light-cone gauge when it approaches $0$.
However when the contribution from the $T$-Wilson line is taken into account, this prescription-dependence cancels and gauge invariance is
completely restored.
\begin{figure}[htp]
% \centering
\vskip-1cm
 \centerline{\includegraphics[width=\columnwidth]{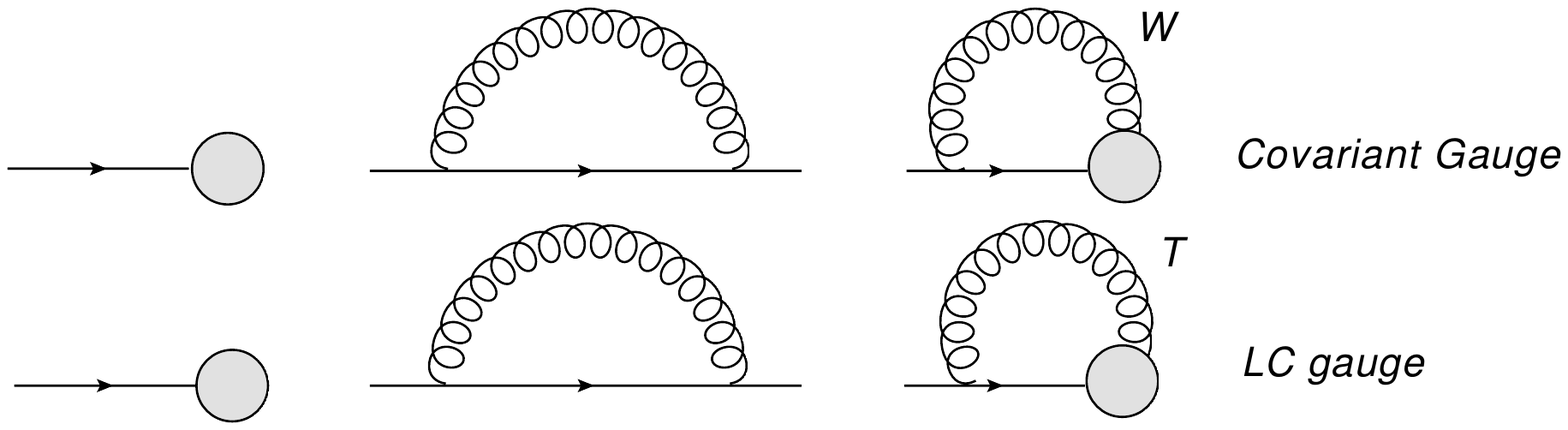}}
 \label{fig:scetformfactor}
\vskip-7cm
\caption{The matrix element for an incoming  jet  at tree level and one loop in SCET in  covariant  and LC gauge}
%\vskip-3cm
 \end{figure}
In Feynman gauge the two diagrams contributing at one-loop level are given in the first line of Figure \ref{fig:scetformfactor}.
We work in dimensional regularization and we take  the external quark momentum $p=(p^+,0,{\bf p_\perp})$ to be off-shell, $p^2=-{\bf p_\perp^2}\neq 0$, $p^+>0$.
Let us denote the contribution from the wave function renormalization graph (WFR) by $\Sigma_{Fey}$ (see e.g. the second reference in ~\cite{SCET1}.)
The contribution from the diagram with collinear Wilson line is given by
\begin{align}
  \label{eq:vFey}
  I_{\bn,Fey}=-2i g^2 C_F \mu^{2\eps}  \int  \frac{d^d k}{(2\pi)^d}\; \frac{1}{(k^2+i0)( k^++i0)}\frac{ p^++ k^+}{(p+k)^2+i0}
  \ ,
\end{align}
   where $C_F=4/3$. The gluon momentum has been chosen such that it is outgoing from the vertex with the collinear Wilson line. The calculation of this integral and others as well are given in the Appendix (see also 
 \cite{cs, Bassetto:1991sv}.)  It is  important to note that in principle this integral contains a  real and an imaginary part.
 The imaginary part of the integral originates from the divergence when $k^+\rightarrow 0$. This imaginary part does not appear if one insists
on using dimensional regularization to regulate IR divergences (see the Appendix).

In light-cone gauge,
the choice of the gauge condition  is $ n\cdot A_{\bn}\equiv A^+_{\bn}=0$ where $n=\frac{1}{\sqrt 2}(0,1,0)$. Clearly this choice renders
$W_{\bn}=1$.
In light-cone gauge and without the contribution from the $T$-Wilson line there is only the WFR graph.
 The gluon propagator in this gauge is given by
\begin{align}
\label{eq:gluonpropLC}
 \tilde D_{\mu\nu}(k)=\frac{-i}{k^2+i0}\left( g_{\mu\nu}-\frac{k_\mu n_\nu+k_\nu n_\mu}{[ k^+]}\right)\ ,
\end{align}
where the symbol $[k^+]$ means that an appropriate prescription condition must be chosen to regularize the spurious light-cone singularity. 
The contribution from $g_{\mu\nu}$ in Eq.~(\ref{eq:gluonpropLC}) to the WFR
  gives the same result as in Feynman gauge. The remaining contribution is known as the ``Axial part''.
 Thus the final result in light-cone gauge is given by
\begin{align}
 \Sigma_{LC}(p)=\Sigma_{Fey}(p)+\Sigma_{Ax}^{({\rm Pres})}(p)=(I_{w,Fey}(p)+I_{w,Ax}^{({\rm Pres})}(p))\frac{i p^2}{p^+}
\frac{\not\!\! n}{\sqrt{2}}\ ,
\end{align}
where the suffix ``Pres'' indicates the prescription dependence of the integral.
The most common prescriptions used in full QCD calculations are listed in Tab.~\ref{tab:prescription}.
\begin{table*}[htbp]
	\centering
		\begin{tabular}{||c|c||}\hline\hline
			Prescription & $1/[k^+]$ \\ \hline
			$+i0$ & $1/(k^++i0)$ \\
			$-i0$ & $1/(k^+-i0)$  \\
			PV &$1/2 (1/(k^++i0)+1/(k^+-i0))$\\
			ML & $ 1/(k^++i0 {\rm Sgn(k^-)})$\\
			\hline\hline
		\end{tabular}
	\caption{The most commonly used prescriptions light-cone gauge.}
	\label{tab:prescription}
\end{table*}
  The value of $I_{w,Fey}(p)$  can be easily extracted from that of  $\Sigma_{Fey}$ and we  do not need it explicitly here.
It is also well-known that $I_{w,Fey}(p)$ is the same  in SCET
and in QCD.  This  is no more true for the axial part, $I_{w,Ax}^{({\rm Pres})}(p)$, whose result of calculation is different in the two cases.
 In SCET  one finds
\begin{align}
\label{eq:sigmaLC}
 \Sigma_{Ax}^{({\rm Pres})}(p)&=I_{w,Ax}^{({\rm Pres})} (p)\frac{\not\!\! n}{\sqrt{2}}\frac{i p^2}{p^+}
\nn \\
&= 4g^2 C_F \mu^{2\eps}
\int \frac{d^d k}{(2\pi)^d}\; \frac{-i}{(k^2+i0)[ k^+]}\left(
\frac{p^+}{p^2}-\frac{ p^++ k^+}{(p+k)^2+i0}\right)\frac{i p^2}{p^+}\frac{\not\!\! n}{\sqrt{2}}\ .
\end{align}
The first integral is identically zero  in all prescriptions that we have studied so we end up with
\begin{align}
\label{eq:IAx}
I_{w,Ax}^{({\rm Pres})}(p)=4ig^2 C_F \mu^{2\eps}
\int \frac{d^d k}{(2\pi)^d}\; \frac{1}{(k^2+i0)[ k^+]}\frac{ p^++ k^+}{(p+k)^2+i0}\ .
\end{align}
The gauge invariance of the SCET matrix element is guaranteed when $I_{\bn,Fey}=(-1/2)I_{Ax}^{(+i0)}$.
 In ML prescription we have
\begin{align}
 \frac{1}{[ k^+]}=\frac{\theta( k^-)}{ k^++i p^+\eta}+\frac{\theta(-k^-)}{ k^+-i p^+ \eta}.
\end{align}
where $\eta>0$ is a  small number and is used to regulate the $k^+ \rightarrow 0$ divergence.
 In the ML prescription one finds  that in Eq.~(\ref{eq:IAx}) only the integrand with $k^+$ in the numerator
 gives a  non-zero result with the off-shellness that we have chosen
   (see e.g.~\cite{Bassetto:1991sv}) and the result is real and independent of $\eta$ however, and most importantly, it has only a single pole. The integral with $p^+$ in the numerator gives zero in ML prescription.
 Subtracting the zero-bin contribution has also no effect as it is zero in the ML prescription (see Appendix) for the off-shellness that we have chosen.
%However while in the PV prescription it is the sum of the two addends
%in Eq.~(\ref{eq:pv})
%that cancels itself, in the ML prescription each of the addends gives a null contribution.
Since the pure jet function in Feynman gauge has a double pole one can easily conclude that the statement that the matrix elements: $\langle 0 \vert W_{\bn}^{\dagger} \xi_{\bn}\vert q \rangle $ and $\langle q|  \bar\xi_nW_n | 0\rangle$
  are gauge invariant in both  type of gauges, regular and singular ones, is in general not true. %prescription-dependent as we stated before.
%It is clear that gauge invariance should be genuinely independent of any prescription.

Let us now consider the matrix element describing an incoming jet
\begin{align}
\label{eq:buildingblocks2}
\langle 0|T_{\bn}^\dagger W_{\bn}^\dagger\xi_\bn|q_\bn\rangle\ ,
\end{align}
where
the $T$-Wilson line is
\begin{align}
 T_{\bn}^\dagger(x^+,x_\perp)=&{\cal P}\exp\left[i g\int_0^\infty d\tau {\bf l}_\perp \cdot {\bf A}_\perp(\infty^-,x^+;{\bf l}_\perp\tau+{\bf x}_\perp)
\right]\ .
 %\nonumber\\
%T^\dagger_n(x^-,x_\perp)=&\overline{{\cal P}}\exp\big[-i g\int_0^\infty d\tau {\bf l}_\perp \cdot {\bf A}_\perp(x^-,\infty^+;
%{\bf l}_\perp\tau+{\bf x}_\perp)
\end{align}
 The  vector $\bf l$ specifies a  path in a two-dimensional space. Notice that in light-cone gauge and at infinity in one light-cone direction,
the gauge field is a gradient of a scalar function thus the final result of integration is path-independent.
In the following we study some of the properties of the new Wilson line $T_\bn^\dagger$ and we show how it fixes gauge invariance issues
discussed earlier.

 First we  note that $ {\bf A}_\perp(\infty^-,x^+; x_\perp)=0$ in covariant gauges.  So $T_\bn^\dagger\equiv 1$ and $\langle 0|T_{\bn}^\dagger
 W_{\bn}^\dagger\xi_\bn|q_\bn\rangle$ reduces to  $\langle 0| W_\bn^\dagger \xi_\bn| q_\bn\rangle$ as it should be.
This implies that all the results obtained in SCET in covariant gauges are still valid even with the introduction of the $T$-Wilson line.
Then we use
\begin{align}
 {\bf A}_\perp(\infty^-,x^+;{\bf l}_\perp \tau)=\int \frac{d k^+}{2\pi} e^{-i [k^+]\infty^-}\int\frac{d k^-}{2\pi} e^{-i ( k^-) x^+}\int
\frac{d k_\perp^2}{(2\pi)^2}
e^{i {\bf k_\perp \cdot l_\perp}\tau} {\bf A_\perp}(k)\ ,
\end{align}
 so that
\begin{align}
\int_0^\infty d\tau {\bf l}_\perp \cdot {\bf A}_\perp(\infty^-,x^+;{\bf l}_\perp\tau)= \hspace{5cm}&\nn \\
\int \frac{d k^+}{2\pi} e^{-i[k^+]\infty^-}
\int\frac{d k^-}{2\pi} e^{-i ( k^-) x^+}
\int\frac{d k_\perp^2}{(2\pi)^2}
%e^{i {\bf k_\perp \cdot l_\perp}\tau}
 \frac{i {\bf l_\perp\cdot A_\perp}(k)} {{\bf k_\perp \cdot l_\perp}-i0}
 \ . &
\end{align}
 The fundamental observation now is that
\begin{align}
\label{eq:cinfty}
 \frac{e^{-i [k^+]\infty^-}}{[k^+]}=\frac{e^{i [k^+]\infty}}{[k^+]}=2\pi i C_\infty^{(\rm Pres)} \delta(k^+)=
 \left[\frac{C_\infty^{(\rm Pres)}}{k^+-i0}-
\frac{C_\infty^{(\rm Pres)}}{k^++i0}\right]\ .
\end{align}
The   $C_\infty^{(\rm Pres)}$ depends on the prescription that is chosen  in light-cone gauge as is shown in
Tab.~(\ref{tab:cinfty}).\footnote{The difference  in $C_\infty^{(\rm Pres)}$ with respect to Ref.~\cite{cs} depend on the fact that in SCET
the Feynman rule for the vertex coming from the $W^\dagger$ Wilson line has a $+i0$, while in Ref.~\cite{cs} they
consider the case of retarded Green functions.}
\begin{table*}[htbp]
	\centering
		\begin{tabular}{||c|c||}\hline\hline
			Prescription & $C_\infty^{(\rm Pres)}$ \\ \hline
			$+i0$ & 0 \\
			$-i0$ & 1 \\
			PV &1/2 \\
			ML & $\theta(-k^-)$\\
			\hline\hline
		\end{tabular}
	\caption{$C_\infty^{(\rm Pres)}$ in  the most used prescriptions for LC gauge}
	\label{tab:cinfty}
\end{table*}

 In the  calculation of loops with $T_\bn^\dagger$ we  use then the  following propagators. When $ \mu$ picks  up collinear indexes, $\mu=+,-$,
and using Eq.~(\ref{eq:cinfty}) one finds
\begin{align}
 \langle A^i_\perp(\infty^-,x^+,x_\perp) A^\mu(0)\rangle&=\lim_{x^-\rightarrow -\infty}
\int \frac{d^4 k}{(2\pi)^4} e^{-i([k^+] x^-+k^- x^++k_\perp x_\perp)}\frac{i}{k^2+i0}\left(
\frac{n^\mu k^i}{[k^+]}\right) \nonumber \\
&=\int \frac{dk^-}{(2\pi)}C_\infty^{(\rm Pres)} e^{-i k^- x^+} \int \frac{d^2 k_\perp}{(2\pi)^2}e^{-i k_\perp x_\perp}\frac{1}{k^2_\perp+i0}
(-n^\mu k^i)
\label{eq:prop1}\nonumber\\
&=i\int \frac{d^4 k}{(2\pi)^4}\frac{e^{-i(k^- x^++k_\perp x_\perp)}(n^\mu k^i)}{k^2+i0}\left[\frac{C_\infty^{(\rm Pres)}}{k^+-i0}-
\frac{C_\infty^{(\rm Pres)}}{k^++i0}\right]\ .
%\label{eq:prop2}
\end{align}
 where the last line of Eq.~(\ref{eq:prop1}) is a useful form  for practical calculations.
When $\mu=j$   the same reasoning as in Eq.~(\ref{eq:prop1}) gives
\begin{align}
 \langle A^i_\perp(\infty^-,x^+,x_\perp) A^j(0)\rangle&=-\delta^{ij}\lim_{x^-\rightarrow -\infty}
\int \frac{d^4 k}{(2\pi)^4} e^{-i([k^+] x^-+k^- x^++k_\perp x_\perp)}\frac{-i}{k^2+i0} \nn \\
&=-\delta^{ij}\lim_{x^-\rightarrow -\infty}
\int \frac{d^4 p}{(2\pi)^4} \frac{e^{-i([k^+] x^-+k^- x^++k_\perp x_\perp)}}{ k^+} \frac{-i k^+}{k^2+i0}\nn \\
&=-\delta^{ij}
\int \frac{d^4 k}{(2\pi)^4} e^{-i(k^- x^++k_\perp x_\perp)} \frac{ k^+ \;\delta( k^+)}{k^2+i0}C_\infty^{(\rm Pres)}=0\ .
\end{align}
Using these rules one can now calculate the one-loop contribution coming from $T_\bn^\dagger$,
\begin{align}
\label{eq:Itax}
 I_{T, Ax}^{(\rm Pres)}&=  2 C_F g^2 \mu^{2\eps}\int \frac{d^d k}{(2\pi)^d}\frac{-l_\perp^\mu}{l_\perp k_\perp} \frac{i(p^++k^+)}{(p+k)^2}
\nn \\ &\times
i
\left[\bn^\nu+\frac{\ga^\nu_\perp \pslash_\perp}{p^+}+ \frac{(\kslash_\perp+ \pslash_\perp)\ga^\nu_\perp }{p^++k^+}-
\frac{(\kslash_\perp+ \pslash_\perp)\pslash_\perp}{p^+(p^++k^+)n^\nu}\right]\nn \\ &\times
\frac{i (n^\nu k^\mu)}{k^2+i0} C_\infty^{(\rm Pres)}  \left[\frac{1}{k^+-i0}-\frac{1}{k^++i0}\right] \nn \\
&= 2 C_F g^2 \mu^{2\eps}i \int \frac{d^d k}{(2\pi)^d}
\frac{ p^++ k^+}{(k^2+i0)((p+k)^2+i0)}\left[\frac{C_\infty^{(\rm Pres)} }{k^+-i0}-\frac{C_\infty^{(\rm Pres)} }{k^++i0}\right]\ .
\end{align}
The last line  of Eq.~(\ref{eq:Itax}) shows explicitly  that the path-dependence of the $T$-Wilson line is canceled. This is to be expected
since at infinity in light-cone direction and in light-cone gauge the gluon field is just a gradient of a scalar function so the line integral is path-independent.
We expect that path independence is a general feature of the $T$-Wilson line, as the cancellation of prescription dependence  is in principle independent of any path. A verification of this statement with the calculation of the matching of the  electromagnetic form factor  between SCET and QCD in light-cone gauge is in progress~\cite{future}.
Summing up the last result with the axial contribution from the wave function renormalization diagram we find that gauge invariance is
 established when
\begin{align}
\label{eq:Igauge}
I_{\bn,Fey}=\frac{-1}{2} I_{w,Ax}^{(\rm ML)}+I_{T, Ax}^{(\rm ML)}\ .
\end{align}
 Using the value of $C_\infty^{(\rm ML)}$ extracted from Eq.~(\ref{eq:cinfty})  and in Tab.~\ref{tab:cinfty},
one finds that  all the prescription-dependence exactly
cancels  in the sum  of the r.h.s. of Eq.~(\ref{eq:Igauge}) and the gauge invariance is realized.
% We observe that in the ``$\pm i0$'' and  in the PV prescriptions gauge invariance is established  just by considering
%the integrands and without the need of any explicit calculation.
%  In the ML prescription $C_\infty^{(\rm Pres)}$ is not a simple constant and the
%so the integrals must be done explicitly. The
% calculation of the relevant  integrals are shown in the Appendix.
Eq.~\ref{eq:Igauge} is actually valid for all other prescriptions as well, 
$I_{\bn,Fey}=- I_{w,Ax}^{(\rm Pres)}/2+I_{T, Ax}^{(\rm Pres)}$, however zero-bin subtraction plays a special role in these other prescriptions as is discussed in the following section.

%%&&&&&&&&&&&&&&&&&&&&&&&&&&&&&&&&&&&&&&&&&&&&&&&&&&&
\section{The SCET jet in $\pm i0$ and PV prescriptions}
%%&&&&&&&&&&&&&&&&&&&&&&&&&&&&&&&&&&&&&&&&&&&&&&&&&&&&&
The light-cone prescriptions $\pm i 0$ and PV are in principle not compatible with the quantization of QCD in light-cone gauge~\cite{Bassetto:1984dq}.
Nevertheless one can ask what happens to the SCET jet  when using these prescriptions in actual calculations.
Comparing the integrands in Eq.~(\ref{eq:vFey}) (see the Appendix) and Eq.~(\ref{eq:IAx}) we see that they agree when the prescription 
$[k^+]= k^++i\eta p^+$ is
chosen. 
To show the difference between the $+i\eta$ and $-i\eta$ prescriptions it is enough at this stage to consider the ultra-violet 
(UV) divergent part of the integral $I_{w,Ax}^{({\rm Pres})}(p)$ where for simplicity of notation we label it as $I_{\bn}^{\rm (Pres)}$
(the calculation of the finite part can be found in the Appendix).%$I_{\pm i\eta}^{\rm UV}$
%. The calculation can proceed by first perform contour integrating over the small component $k^-$, then by dimensional regularization in $d=2-2\varepsilon$ over $|{\vec k}|$.
 The result is,
\begin{eqnarray}
%I_{\pm i\eta}^{\rm UV}
\label{eta100}
I_{\bn}^{\pm i\eta}
=-2\frac{\alpha_s}{4\pi}C_F\left(\frac{\mu^2}{-p^2}\right)^{\varepsilon}\frac{1}{\varepsilon}\; \int_0^1 dz \frac{(1-z)^{1-\varepsilon}z^{-\varepsilon}}{z\mp i\eta}\  .
\end{eqnarray}
The singularity at $z=0$ is regularized by the contour deformation. In this sense the $\pm i\eta$ serves to regularize the IR collinear divergence and thus we can Taylor expand the numerator in powers of
$\varepsilon$. Keeping the leading term one gets finally for the divergent part
\begin{eqnarray}
I_{\bn}^{\pm i\eta}=-\frac{g^2}{4\pi^2}C_F\frac{1}{\varepsilon}\left[1+\ln \mp i\eta \right]\
\end{eqnarray}
and with a Principal Value (PV) prescription,
\begin{align}
\label{eq:pv}
{\rm Im} I_{\bn}^{(PV)}(p)=\frac{1}{2}({\rm Im}I_{\bn}^{(+i0)}(p)+{\rm Im}I_{\bn}^{(-i0)}(p) )=0 \ .
\end{align}
Thus all these prescriptions give results which depend on $\eta$ and the imaginary part of the integral varies with the
prescription.
If one  regulates the IR collinear
singularity within pure dimensional regularization %(which is equivalent to setting $\eta=0$ in Eq.~(\ref{eta100})) 
 both ${\rm Im}I_{\bn}^{(+i0)}(p)=0$ and ${\rm Im}I_{\bn}^{(-i0)}(p)=0$ so that it seems
 that gauge invariance is achieved (also without the introduction of $T$). However  this is a particular feature of an IR regulator.
In principle the  insertion of $T$ and the use of the equivalent of Eq.~\ref{eq:Igauge},
$I_{\bn,Fey}=- I_{w,Ax}^{(\rm Pres)}/2+I_{T, Ax}^{(\rm Pres)}$, and  $C_\infty^{(\rm Pres)}$
as given in Tab.~\ref{tab:cinfty}, removes completely the gauge changing imaginary part of the integrals. The $T$ so restores gauge invariance
with all IR regulators.

Now we consider the pure collinear part of the matrix element, i.e., 
the one after the zero-bin contribution is subtracted.  
The fundamental observation is that for $\pm i0$ and PV prescriptions, the zero-bin subtraction cancels completely 
all the dependence on the IR parameter $\pm i \eta$ leaving us with the result of Feynman gauge (see the Appendix for the zero-bin contributions). 
In other words the zero-bin is responsible of canceling all prescription and IR-regulator dependence in the case of the $\pm i0$, PV prescriptions. 
Thus we see that at least at one loop one does not really need to introduce the $T$-Wilson line in those prescriptions
(altough it is certainly possible to define a matrix element and its zero-bin subtraction both  and separately gauge invariant with 
the use of $T$). 
However as we established in the previous section this is not the case in the ML prescription which is 
the only  one  reliably consistent with the quantization of  QCD in light-cone gauge.
  
%The above observation however can be recast in the following way.  We can ask as in Ref.~\cite{cs, ji} that
%both matrix elements, the naive collinear one (i.e., without zero-bin subtraction) and the zero-bin  matrix element (in the sense of ``soft function'' as discussed in \cite{Lee:2006nr,Idilbi:2007ff,Idilbi:2007ff}) are separately
%gauge invariant. This is achieved only by considering the  contribution of the $T$-Wilson line in both matrix elements.
%In fact  the role of the $T$  is that it removes the imaginary (prescription-dependent) part of the collinear integrals while
%the zero-bin subtraction finally removes the rest of the dependence on the infrared  regulator $\eta$.

%We observe that the r.h.s. in  Eq.~(\ref{eq:cinfty}) is exactly 0 in the ``$+i0$'' prescription thus $T_\bn^\dagger=1$.
%This is consistent with our calculation of  $\langle 0| W_\bn^\dagger \xi_\bn| q_\bn\rangle$ where we have seen in light-cone gauge with  ``$+i0$'' the result is the same as in Feynman gauge and no transverse gauge link is needed.

%%&&&&&&&&&&&&&&&&&&&&&&&&&&&&&&&&&&&&&&&&&&&&&&&&
\section{Applications and conclusions}
%%&&&&&&&&&&&&&&&&&&&&&&&&&&&&&&&&&&&&&&&&&&&&&&&&&&&&&&&
Until now we have considered the contribution from the $T$-Wilson line in one collinear direction however in many applications of SCET
there are more than one collinear direction. To make the discussion more concrete let us consider the quark form factor both in full QCD and in SCET.
 In full QCD it is well-know that the electromagnetic current $j^\mu=\bar {\psi}\gamma^\mu \psi$ is a conserved quantity and is gauge invariant under
arbitrary $SU(3)$ gauge transformation. In SCET it has been established by different authors~\cite{varia} that the full QCD quark form factor
$\langle q(p_2)\vert j^\mu \vert q(p_1)\rangle$ factorizes into an incoming jet, outgoing jet and a soft function built out of two soft Wilson lines.
However all those treatments were actually performed in Feynman gauge and not in light-cone gauge.
The calculation in SCET in light-cone gauge can now be performed with the $T$-Wilson line that enters into both jets and one needs to introduce
two light-cone gauge fixing conditions for each collinear jet (or collinear Lagrangian) and two different $T$ Wilson lines again for each collinear
 direction, say $T_n$ and $T_{\bn}$ (and/or their Hermitian conjugates.) With these two transverse Wilson lines each of the two jets becomes gauge
invariant in both regular and singular gauges. Thus using $T$'s the factorization of the full QCD quark form factor established
in covariant gauges carries through  straightforwardly also in light-cone gauge.

As we mentioned before the practical importance of the $T$-Wilson lines in SCET is that it allows us to write  a
gauge invariant definitions of the non-perturbative matrix elements appearing in the factorization theorems for certain cross-sections
where fields are separated in the transverse direction.

As an example we take the TMDPDF in QCD and in SCET.
 In QCD the  gauge invariant definition  was first given in Ref.~\cite{yuan} and studied in light-cone gauge in Ref.~\cite{cs}.
In all these works it is pointed out how in QCD new kind of divergences appear
when the fields entering  the matrix elements are separated in the transverse direction.
The TMDPDF occur for instance in SIDIS  and here the complete factorization is achieved only in the presence of transverse links.
 In particular in Ref.~\cite{cs} it is  shown  how to use light-cone gauge  for practical calculations in SIDIS.
Within SCET  the TMDPDF  for a quark $q$ in a hadron $P$ with momentum $p$ can be defined with the use of   $\underline{\chi}$-field
\begin{align}
\label{eq:tmdpdf}
\underline{\chi}_\bn (y)&\equiv T^\dagger_\bn (y^+,{\bf y}_\perp)W^\dagger_\bn(y) \xi_\bn (y)  , \nonumber \\
\phi_{q/P}&=\langle P_\bn|\overline{\underline{\chi}}_\bn (y)\delta\left(x-\frac{n {\cal P}}{n p}\right)\delta^{(2)}(p_\perp-{\cal P}_\perp)
\frac{\nslash}{\sqrt{2}}
\underline{\chi}_\bn (0)|P_\bn\rangle \ ,
\end{align}
where $x$ is the momentum fraction of the quark in the light-cone direction, and ${\cal P}$ is the usual label operators in SCET.
The above definition of the TMDPDF is now gauge invariant in both classes of gauges.
 The complete analysis of TMDPDF in SCET and its application to SIDIS and other  semi-inclusive processes will be developed  in a forthcoming
work~\cite{future}.

The above analysis of the TMDPDF can be straightforwardly extended to consider the non-perturbative matrix elements introduced
in ~\cite{Mantry:2009qz,D'Eramo:2010ak,Becher:2010tm,Becher:2010pd} where one needs to invoke the $T$-Wilson line in the operator definition
to obtain gauge invariant matrix elements as they should be.
A final comment concerns the use  of prescriptions in QCD. In ref.~\cite{Bassetto:1984dq} it was argued that the only prescription 
 that allows the quantization of QCD is the ML prescription.  However  in SCET  we are considering just the Fock space of collinear particles
and not the whole Fock space of QCD.
The  fact that we are considering a restriction of the whole  Fock space implies that we  have to re-discuss the quantization of the theory
in this particular subspace~\cite{AIIS} with the help of SCET. Moreover in Ref.~\cite{Bassetto:1984dq} the role of the $T$-Wilson line was not discussed. In Ref.~\cite{Becher:2010pd} the authors  use SCET in light-cone gauge and find that their calculation cannot
 give the correct  result in the ML prescription (so contradicting the results of Ref.~\cite{Bassetto:1984dq}).
 We think that  the inclusion of the $T$-Wilson line would  reconcile the calculation in Ref.~\cite{Becher:2010pd}
with the ML prescription~\cite{AIIS}.

To conclude we have shown that the introduction of the $T$-Wilson line in SCET is mandatory to achieve gauge invariant matrix elements both in covariant as well as in singular gauges. We explicitly performed one loop calculation for the SCET fundamental quantity $\langle 0| W_\bn^\dagger \xi_\bn| q_\bn\rangle$ in both Feynman and light-cone gauges and showed that in the ML prescription the $T$-Wilson line must be introduced to obtain gauge invariance. In other prescriptions commonly used in light-cone gauge, the soft or zero-bin subtractions remove the prescription dependence thus the purely collinear SCET matrix elements are gauge invariant in those prescriptions. However as we mentioned earlier such prescriptions may not lead to a proper quantization of SCET as it is the case in QCD.

The introduction of the $T$-Wilson line allows SCET to properly define in a gauge invariant way non-perturbative matrix elements such as the TMDPDF and generalizations thereof.

\section*{ACKNOWLEDGMENTS}
This work was supported by the EU network, MRTN-CT-2006-035482 (Flavianet),
Spanish MEC, FPA2008-00592 and Banco Santander UCM-BSCH GR58/08 910309. I.S.  is supported by Ramon y Cajal Program. A.I.  is supported
 by the Spanish grant CPAN-ingenio 2010. We enjoyed discussions with Miguel Garc\'ia Echevarr\'ia and Iain W. Stewart.
\section*{Appendix}

Considering the axial part of the WFR diagram the integral we  need to calculate in $\pm i 0$ prescriptions  is
\begin{align}
  \label{eq:vFey2}
  I_{\bn}^{\pm i 0}=-2i g^2 C_F \mu^{2\eps}  \int  \frac{d^d k}{(2\pi)^d}\; \frac{1}{(k^2+i0)( k^+\pm i0)}\frac{ p^++ k^+}{(p+k)^2+i0}
  \ ,
\end{align}
 We regularize the IR divergence  with  ($\eta>0$)
\begin{align}
  I_{\bn}^{\pm i 0}&=-2i g^2 C_F \mu^{2\eps}  \int  \frac{d^d k}{(2\pi)^d}\; \frac{1}{(k^2+i0)( k^+\pm ip^+ \eta)}\frac{ p^++ k^+}{(p+k)^2+i0}
  \  \nn  \\
&= -2 \frac{\alpha_s}{4 \pi} C_F \left(\frac{-p^2}{\mu^2}\right)^{-\eps} \Gamma (\eps) \int_0^1 dz\;  z^{-\eps} (1-z)^{1-\eps}
  \frac{1}{z\mp i \eta}
\nn \\
&=-2 \frac{\alpha_s}{4 \pi} C_F  \left\{\frac{1}{\eps}
\left(
-1-\ln (\mp i \eta)\right)
+\ln \left(\frac{-p^2}{\mu^2}\right)(1+ \ln (\mp i \eta))+\frac{1}{2}\ln^2(\mp i \eta)-2+\frac{\pi^2}{3}
\right\}
\end{align}
This result is in agreement  with Ref.~\cite{cs} when $\eta$ is small.
The zero-bin part of this integral, that must be subtracted is
\begin{align}
 I_{\bn,0}^{\pm i 0}&=-2i g^2 C_F \mu^{2\eps}  \int  \frac{d^d k}{(2\pi)^d}\;
\frac{1}{(k^2+i0)( k^+\pm ip^+ \eta)}\frac{ p^+}{p^2+k^- p^++i0}
  \  \nn  \\
&=-2 \frac{\alpha_s}{4 \pi} C_F \left\{\frac{1}{\eps^2}-\frac{1}{\eps}\left(\ln (\mp i \eta)+\ln \left(\frac{-p^2}{\mu^2}\right)\right)
+\frac{1}{2}\left( \ln (\mp i \eta) +\ln \left(\frac{-p^2}{\mu^2}\right)\right)^2+\frac{\pi^2}{4}
\right\}
\end{align}
 Subtracting $I_{\bn,\pm i 0}-I_{\bn,0}^{\pm i 0}$ one recovers the result of A. Manohar in Ref.~\cite{varia}.
In the ML prescription, with the off-shellness that we have considered we have
\begin{align}
\label{MLi}
 \mu^{2\eps}  \int  \frac{d^d k}{(2\pi)^d}\; \frac{1}{(k^2+i0)( k^++ ip^+ \eta Sgn(k^-))}\frac{ p^+}{(p+k)^2+i0}=0.
\end{align}
For a  general off-shellness the result can be  found in ref.~\cite{Bassetto:1991sv}.
Also the corresponding  zero-bin  subtracted integral, with the off-shellness that we have chosen
\begin{align}
\label{ML0}
 \int  \frac{d^d k}{(2\pi)^d}\;
\frac{1}{(k^2+i0)( k^++ ip^+ \eta Sgn(k^-))}\frac{ p^+}{p^2+k^- p^++i0}=0.
\end{align}
When integrating in the  complex $k^+$-plane  all poles in the integrand of Eq.~(\ref{MLi}) and Eq.~(\ref{ML0}) lie on the
 same side so the integrals  give null results.


\begin{thebibliography}{99}
\bibitem{SCET1}
%\cite{Bauer:2000ew}
%\bibitem{Bauer:2000ew}
  C.~W.~Bauer, S.~Fleming and M.~E.~Luke,
%``Summing Sudakov logarithms in B --> X/s gamma in effective field  theory,''
  Phys.\ Rev.\  D {\bf 63}, 014006 (2000)
  [arXiv:hep-ph/0005275];
%%CITATION = PHRVA,D63,014006;%%
%
%\cite{Bauer:2000yr}
%\bibitem{Bauer:2000yr}
  C.~W.~Bauer, S.~Fleming, D.~Pirjol and I.~W.~Stewart,
  %``An effective field theory for collinear and soft gluons: Heavy to light
  %decays,''
  Phys.\ Rev.\  D {\bf 63} (2001) 114020
  [arXiv:hep-ph/0011336].
  %%CITATION = PHRVA,D63,114020;%%

\bibitem{SCETf}
%\cite{Bauer:2001yt}
%\bibitem{Bauer:2001yt}
  C.~W.~Bauer, D.~Pirjol and I.~W.~Stewart,
  %``Soft-Collinear Factorization in Effective Field Theory,''
  Phys.\ Rev.\  D {\bf 65} (2002) 054022
  [arXiv:hep-ph/0109045].
  %%CITATION = PHRVA,D65,054022;%%

%\cite{Bauer:2002nz}
\bibitem{Bauer:2002nz}
  C.~W.~Bauer, S.~Fleming, D.~Pirjol, I.~Z.~Rothstein and I.~W.~Stewart,
  %``Hard scattering factorization from effective field theory,''
  Phys.\ Rev.\  D {\bf 66}, 014017 (2002)
  [arXiv:hep-ph/0202088];
  %%CITATION = PHRVA,D66,014017;%%
%
%\cite{Beneke:2002ph}
%\bibitem{Beneke:2002ph}
  M.~Beneke, A.~P.~Chapovsky, M.~Diehl and T.~Feldmann,
  %``Soft-collinear effective theory and heavy-to-light currents beyond  leading
  %power,''
  Nucl.\ Phys.\  B {\bf 643}, 431 (2002)
  [arXiv:hep-ph/0206152].
  %%CITATION = NUPHA,B643,431;%%

  \bibitem{jack}
%\cite{Jackiw:1991ck}
%\bibitem{Jackiw:1991ck}
  R.~Jackiw, D.~N.~Kabat and M.~Ortiz,
  %``Electromagnetic fields of a massless particle and the eikonal,''
  Phys.\ Lett.\  B {\bf 277} (1992) 148
  [arXiv:hep-th/9112020];
  %%CITATION = PHLTA,B277,148;%%
%
 I.~Robinson and K.~Rozga, J.\ Math.\ Phys.\ {\bf 25}, 499 (1984);
%\cite{'tHooft:1987rb}
%\bibitem{'tHooft:1987rb}
  G.~'t Hooft,
  %``Graviton Dominance in Ultrahigh-Energy Scattering,''
  Phys.\ Lett.\  B {\bf 198} (1987) 61.
  %%CITATION = PHLTA,B198,61;%%

\bibitem{ji}
%\cite{Ji:2004wu}
%\bibitem{Ji:2004wu}
  X.~d.~Ji, J.~p.~Ma and F.~Yuan,
  %``QCD factorization for semi-inclusive deep-inelastic scattering at low
  %transverse momentum,''
  Phys.\ Rev.\  D {\bf 71} (2005) 034005
  [arXiv:hep-ph/0404183];
  %%CITATION = PHRVA,D71,034005;%%
%\cite{Collins:2000gd}
%\bibitem{Collins:2000gd}
  J.~C.~Collins and F.~Hautmann,
  %``Soft gluons and gauge-invariant subtractions in NLO parton-shower Monte
  %Carlo event generators,''
  JHEP {\bf 0103} (2001) 016
  [arXiv:hep-ph/0009286];
  %%CITATION = JHEPA,0103,016;%%
%
%\cite{Collins:1999dz}
%\bibitem{Collins:1999dz}
 % J.~C.~Collins and F.~Hautmann,
  %``Infrared divergences and non-lightlike eikonal lines in Sudakov
  %processes,''
  Phys.\ Lett.\  B {\bf 472} (2000) 129
  [arXiv:hep-ph/9908467];
  %%CITATION = PHLTA,B472,129;%%
%
%\cite{Collins:1992kk}
%\bibitem{Collins:1992kk}
  J.~C.~Collins,
  %``Fragmentation of transversely polarized quarks probed in transverse
  %momentum distributions,''
  Nucl.\ Phys.\  B {\bf 396}, 161 (1993)
  [arXiv:hep-ph/9208213].
  %%CITATION = NUPHA,B396,161;%%
\bibitem{cs}
%\cite{Cherednikov:2007tw}
%\bibitem{Cherednikov:2007tw}
  I.~O.~Cherednikov and N.~G.~Stefanis,
  %``Renormalization, Wilson lines, and transverse-momentum dependent parton
  %distribution functions,''
  Phys.\ Rev.\  D {\bf 77} (2008) 094001
  [arXiv:0710.1955 [hep-ph]];
  %%CITATION = PHRVA,D77,094001;%%
%\cite{Cherednikov:2008ua}
%
%\bibitem{Cherednikov:2008ua}
 % I.~O.~Cherednikov and N.~G.~Stefanis,
  %``Wilson lines and transverse-momentum dependent parton distribution
  %functions: A renormalization-group analysis,''
  Nucl.\ Phys.\  B {\bf 802} (2008) 146
  [arXiv:0802.2821 [hep-ph]];
  %%CITATION = NUPHA,B802,146;%%
%
%\cite{Cherednikov:2009wk}
%\bibitem{Cherednikov:2009wk}
 % I.~O.~Cherednikov and N.~G.~Stefanis,
  %``Renormalization-group properties of transverse-momentum dependent parton
  %distribution functions in the light-cone gauge with the Mandelstam-Leibbrandt
  %prescription,''
  Phys.\ Rev.\  D {\bf 80} (2009) 054008
  [arXiv:0904.2727 [hep-ph]];
  %%CITATION = PHRVA,D80,054008;%%
%
%\cite{Hautmann:2007uw}
%\bibitem{Hautmann:2007uw}
  F.~Hautmann,
  %``Endpoint singularities in unintegrated parton distributions,''
  Phys.\ Lett.\  B {\bf 655} (2007) 26
  [arXiv:hep-ph/0702196].
  %%CITATION = PHLTA,B655,26;%%
%
\bibitem{future}
 M. Garc\'ia Echevarr\'ia, A. Idilbi and I. Scimemi, in preparation.
\bibitem{Manohar:2006nz}
  A.~V.~Manohar and I.~W.~Stewart,
  %``The zero-bin and mode factorization in quantum field theory,''
  Phys.\ Rev.\  D {\bf 76}, 074002 (2007)
  [arXiv:hep-ph/0605001].
  %%CITATION = PHRVA,D76,074002;%%
%\cite{Bassetto:1984dq}

\bibitem{Bassetto:1984dq}
  A.~Bassetto, M.~Dalbosco, I.~Lazzizzera and R.~Soldati,
  %``Yang-Mills Theories In The Light Cone Gauge,''
  Phys.\ Rev.\  D {\bf 31} (1985) 2012.
  %%CITATION = PHRVA,D31,2012;%%
%\cite{Bassetto:1991sv}
\bibitem{Bassetto:1991sv}
  A.~Bassetto,
  ``The Light cone Feynman rules beyond tree level'', DFPD-91-TH-19, Sep 1991
  %%CITATION = DFPD-91-TH-19;%%
%\cite{Idilbi:2007ff}






\bibitem{varia}
%\cite{Manohar:2003vb}
A.~V.~Manohar,
 Phys.\ Rev.\ D {\bf 68}, 114019 (2003)
 [arXiv:hep-ph/0309176];
%%CITATION = PHRVA,D68,114019;%%
%


%%%\bibitem{Idilbi:2007ff}
%%%  A.~Idilbi and T.~Mehen,
  %``On the equivalence of soft and zero-bin subtractions,''
%%%  Phys.\ Rev.\  D {\bf 75}, 114017 (2007)
%%%  [arXiv:hep-ph/0702022].
  %%CITATION = PHRVA,D75,114017;%%
%\cite{Lee:2006nr}
%%%\bibitem{Lee:2006nr}
%%%  C.~Lee and G.~F.~Sterman,
  %``Momentum flow correlations from event shapes: Factorized soft gluons and
  %soft-collinear effective theory,''
%%%  Phys.\ Rev.\  D {\bf 75}, 014022 (2007)
%%%  [arXiv:hep-ph/0611061].
  %%CITATION = PHRVA,D75,014022;%%
%\cite{Idilbi:2007ff}
%%%A.~Idilbi and T.~Mehen,
%%%Phys.\ Rev.\ D {\bf 75}, 114017 (2007)
%%%[arXiv:hep-ph/0702022];
%%CITATION = PHRVA,D75,114017;%%
%
%\cite{Chay:2005rz}
%%%J.~Chay and C.~Kim,
%%%Phys.\ Rev.\ D {\bf 75}, 016003 (2007)
%%%[arXiv:hep-ph/0511066].
%%CITATION = PHRVA,D75,016003\textsl{};%%
%
\bibitem{yuan}
%\cite{Ji:2002aa}
%\bibitem{Ji:2002aa}
  X.~d.~Ji and F.~Yuan,
  %``Parton distributions in light-cone gauge: Where are the final-state
  %interactions?,''
  Phys.\ Lett.\  B {\bf 543}, 66 (2002)
  [arXiv:hep-ph/0206057].
  %%CITATION = PHLTA,B543,66;%%
%\cite{Mantry:2009qz}
\bibitem{Mantry:2009qz}
  S.~Mantry and F.~Petriello,
  %``Factorization and Resummation of Higgs Boson Differential Distributions in
  %Soft-Collinear Effective Theory,''
  Phys.\ Rev.\  D {\bf 81} (2010) 093007
  [arXiv:0911.4135 [hep-ph]].
  %%CITATION = PHRVA,D81,093007;%%

%\cite{D'Eramo:2010ak}
\bibitem{D'Eramo:2010ak}
  F.~D'Eramo, H.~Liu and K.~Rajagopal,
  %``Transverse Momentum Broadening and the Jet Quenching Parameter, Redux,''
  arXiv:1006.1367 [hep-ph].
  %%CITATION = ARXIV:1006.1367;%%

%\cite{Becher:2010tm}
\bibitem{Becher:2010tm}
  T.~Becher and M.~Neubert,
  %``Drell-Yan production at small q_T, generalized parton distributions and the
  %collinear anomaly,''
  arXiv:1007.4005 [hep-ph];
  %%CITATION = ARXIV:1007.4005;%%
%
%\cite{Becher:2010pd}
\bibitem{Becher:2010pd}
  T.~Becher and G.~Bell,
  %``The gluon jet function at two-loop order,''
  arXiv:1008.1936 [hep-ph].
  %%CITATION = ARXIV:1008.1936;%%

\bibitem{AIIS}
A. Idilbi and I. Scimemi, in preparation.


\end{thebibliography}
\end{document}